\documentclass[twocolumn]{svjour3}
\usepackage{times,natbib}
\usepackage{graphicx}
\usepackage{psfrag}
\usepackage{url}
\usepackage{amsmath}
\usepackage{amssymb}
\usepackage{color}
\usepackage{lscape}
\usepackage{float}
\usepackage{caption}
\usepackage{rotating}
\usepackage{appendix}
\usepackage{soul}

\newcommand{\bTheta}{\boldsymbol{\Theta}}
\newcommand{\bV}{\boldsymbol{V}}
\newcommand{\bU}{\boldsymbol{U}}
\newcommand{\bD}{\boldsymbol{D}}
\newcommand{\bX}{\boldsymbol{X}}
\newcommand{\bY}{\boldsymbol{Y}}
\newcommand{\bW}{\boldsymbol{W}}
\newcommand{\bZ}{\boldsymbol{Z}}

\begin{document}

\title{Sampling from Dirichlet process mixture models with unknown concentration
parameter: Mixing issues in large data implementations}
\author{David I. Hastie$^\ast$ \and Silvia Liverani$^\ast$ \and Sylvia Richardson}

\institute{David I. Hastie \at
              Imperial College London, UK
\and
Silvia Liverani \at
              Imperial College London, UK and \\
	MRC Biostatistics Unit, Cambridge, UK \and
Sylvia Richardson\at
	MRC Biostatistics Unit, Cambridge, UK \\
              \email{sylvia.richardson@mrc-bsu.cam.ac.uk}  \and
$^\ast$ Joint first authors
}

\date{Received: date / Accepted: date}

\maketitle

\begin{abstract}
We consider the question of Markov chain Monte Carlo sampling from a general stick-breaking Dirichlet process mixture model, with concentration parameter $\alpha$. This paper introduces a Gibbs sampling algorithm that combines the slice sampling approach of \cite{W07} and the retrospective sampling approach of \cite{PR08}. Our general algorithm is implemented as efficient open source C++ software, available as an R package, and is based on a blocking strategy similar to that suggested by \cite{P08} and implemented by \cite{YPR11}.

We discuss the difficulties of achieving good mixing in MCMC samplers of this nature in large data sets and investigate sensitivity to initialisation. We additionally consider the challenges when an additional layer of hierarchy is added such that joint inference is to be made on $\alpha$. We introduce a new label-switching move and compute the marginal partition posterior to help to surmount these difficulties. Our work is illustrated using a profile regression \citep{MPJ10} application, where we demonstrate good mixing behaviour for both synthetic and real examples.
\end{abstract}

\section{Introduction}
\label{sec:intro}
Fitting mixture distributions to model some observed data is a common inferential strategy within statistical modelling, used in applications ranging from density estimation to regression analysis. Often, the aim is not only to fit the mixture, but additionally to use the fit to guide future predictions. Approaching the task of mixture fitting from a parametric perspective, the task to accomplish is to cluster the observed data and (perhaps simultaneously) determine the cluster parameters for each mixture component. This task is significantly complicated by the need to determine the number of mixture components that should be fitted, typically requiring complicated Markov chain Monte Carlo (MCMC) methods such as reversible jump MCMC techniques \citep{RG97} or related approaches involving parallel tempering methods \citep{JHS05}.

An increasingly popular alternative approach to parametric modelling is to adopt a Bayesian non-parametric approach, fitting an infinite mixture, thereby avoiding determination of the number of clusters. The Dirichlet process \citep{F73} is a well studied stochastic process that is widely used in Bayesian non-parametric modelling, with particular applicability for mixture modelling. The use of the Dirichlet process in the context of mixture modelling is the basis of this paper and we shall refer to the underlying model as the Dirichlet process mixture model, or DPMM for brevity.

The idea of sampling from the DPMM is not new and has been considered by a number of authors including \cite{EW95}, \cite{N00}, \cite{IJ01}, and \cite{YPR11}. While the continual evolution of samplers might implicitly suggest potential shortcomings of previous samplers, new methods are often illustrated on synthetic or low dimensional datasets which can mask issues that might arise when using the method on problems of even modest dimension. In fact, it appears that little explicit discussion has been presented detailing the inherent difficulties of using a Gibbs (or Metropolis-within-Gibbs) sampling approach to update such a complex model space, although there are some exceptions, for example \cite{JN07}, in the context of adding additional split-merge type moves into their sampler.

For real (rather than synthetic) data applications of the DPMM, the state space can be highly multimodal, with well separated regions of high posterior probability co-existing, often corresponding to clusterings with different number of components. We demonstrate that such highly multimodal spaces present difficulties for the existing sampling methods to escape the local modes, with poor mixing resulting in inference that is influenced by sampler initialisation. In the most serious case, this can be interpreted as non-convergence of the MCMC sampler. A primary contribution of this paper is to demonstrate these issues, highlighting that if only certain marginals are used to determine convergence they may fail to identify any issue. To address this we introduce the \emph{Marginal Partition Posterior} as a more robust way of monitoring convergence.

A secondary (and more subtle) mixing issue relates to the mixing across the ordering of clusters in a particular clustering process, when a stick breaking construction is used. As we shall detail, such issues are particularly important when simultaneous inference is desired for the concentration parameter $\alpha$, as defined in the following section. This mixing issue was highlighted by \cite{PR08} who observed that the inclusion of label-switching moves can help to resolve the problem. We demonstrate that the moves that they propose offer only a partial solution to the problem, and we suggest an additional label-switching move that appears to enhance the performance of our own implementation of a DPMM sampler.

In the following section, we present the further details of the DPMM. Section \ref{sec:mixing} discusses some of the mixing issues with DPMM samplers, including Section \ref{sec:label} where we introduce the new label-switching move. This is followed by Section \ref{sec:mpp} where we present a method that we have found useful for determining sampler convergence. The implementation of our sampler is briefly summarised in Section \ref{sec:implement} before Section \ref{sec:application} demonstrates some of the earlier ideas in the context of a real data example.

\section{Dirichlet process mixture models}
\label{sec:dpmm}
A variety of ways have been used to show the existence of the Dirichlet Process, using a number of different formulations \citep{F73,BM73}. In this paper we focus on Dirichlet process mixture models (DPMM), based upon the following constructive definition of the Dirichlet process, due to \cite{S94}. If
\begin{eqnarray}
 \nonumber
 P & = & \sum_{c=1}^{\infty}\psi_c\delta_{\Theta_c},\\
 \nonumber
 \Theta_c &\sim& P_{\Theta_0} \;\;\textrm{for }c\in\mathbb{Z}^{+},\\
 \label{eqn:stick}
 \psi_c & = & V_c\prod_{l<c}(1-V_l)\;\;\textrm{for
}c\in\mathbb{Z}^{+}\setminus\{1\},\\
 \nonumber
 \psi_1 & = & V_1,\;\;\textrm{ and}\\
 \nonumber
 V_c & \sim & \mathrm{Beta}(1,\alpha)\;\;\textrm{for }c\in\mathbb{Z}^{+},
\end{eqnarray}
where $\delta_x$ denotes the Dirac delta function concentrated at $x$, then $P~\sim\mathrm{DP}(\alpha,P_{\Theta_0})$. This formulation for $\bV$ and $\boldsymbol{\psi}$ is known as a \emph{stick-breaking} distribution. Importantly, the distribution $P$ is discrete, because draws $\tilde{\Theta}_1,\tilde{\Theta}_2,\ldots$ from $P$ can only take the values in the set $\{\Theta_c:c\in\mathbb{Z}^{+}\}$.

It is possible to extend the above formulation to more general stick-breaking formulations \citep{IJ01,KGW11,PY97}.

\subsection{Sampling from the DPMM}
For the DPMM, the (possibly multivariate) observed data $\bD=(D_1,D_2,\ldots,D_n)$ follow an infinite mixture distribution, where component $c$ of the mixture is a parametric density of the form $f_c(\cdot)=f(\cdot|\Theta_c,\Lambda)$ parametrised by some component specific parameter $\Theta_c$ and some global parameter $\Lambda$. Defining (latent) parameters $\tilde{\Theta}_1,\tilde{\Theta}_2,\ldots,\tilde{\Theta}_n$ as draws from a probability distribution $P$ following a Dirichlet process $DP(\alpha,P_{\Theta_0})$ and again denoting the dirac delta function by $\delta$, this system can be written,
\begin{eqnarray}
 \label{eqn:dpmm1}
  D_i|\tilde{\Theta}_i,\Lambda &\sim
&f(D_i|\tilde{\Theta}_i,\Lambda)\;\;\textrm{for
}i=1,2,\ldots,n,\\
 \nonumber
 \tilde{\Theta}_i & \sim & \sum_{c=1}^{\infty}\psi_c\delta_{\Theta_c}
\;\;\textrm{for }i=1,2,\ldots,n.
\end{eqnarray}

When making inference using mixture models (either finite or infinite) it is common practice to introduce a vector of latent allocation variables $\bZ$. Such variables enable us to explicitly characterise the clustering and additionally facilitate the design of MCMC samplers. Adopting this approach and writing $\boldsymbol{\psi}=(\psi_1,\psi_2,\ldots)$ and $\bTheta=(\Theta_1,\Theta_2,\ldots)$, we re-write Equation \ref{eqn:dpmm1} as
\begin{eqnarray}
\nonumber
  D_i|\bZ,\bTheta,\Lambda &\sim
&f(D_i|\Theta_{Z_i},\Lambda)\;\;\textrm{for }i=1,2,\ldots,n,\\
 \nonumber
 \Theta_c &\sim& P_{\Theta_0} \;\;\textrm{for }c\in\mathbb{Z}^{+},\\
 \label{eqn:dpmm2}
 \mathbb{P}(Z_i=c|\boldsymbol{\psi}) & = & \psi_c \;\;\textrm{for
}c\in\mathbb{Z}^{+},\;i=1,2,\ldots,n.
\end{eqnarray}
We refer to the model in Equation \ref{eqn:dpmm2}, with no variables integrated out, as the \emph{full stick-breaking DPMM} or even the \emph{FSBDPMM} for conciseness.

Historically, methods to sample from the DPMM \citep{EW95,N00} have simplified the sample space of the full stick-breaking DPMM by integrating out the mixture weights $\boldsymbol{\psi}$. Collectively, such samplers have been termed \emph{Poly\`a Urn} samplers. \cite{IJ01} presented a number of methods for extending Poly\`{a} Urn samplers, and additionally suggested a truncation approach for sampling from the full stick-breaking DPMM with no variables integrated out.

More recently, two alternative innovative approaches to sample directly from the FSBDPMM have been proposed. The first, introduced by \cite{W07} and generalised by \cite{KGW11}, uses a novel slice sampling approach, resulting in full conditionals that may be explored by the use of a Gibbs sampler. The second distinct MCMC sampling approach was proposed in parallel by \cite{PR08}. The proposed sampler again uses a Gibbs sampling approach, but is based upon an idea termed \emph{retrospective sampling}, allowing a dynamic approach to the determination of the number of components (and their parameters) that adapts as the sampler progresses. The cost of this approach is an ingenious but complex Metropolis-within-Gibbs step, to determine cluster membership. Despite the apparent differences between the two strategies, \cite{P08} noted that the two algorithms can be effectively combined to yield an algorithm that improves either of the originals. The resulting sampler was implemented and presented by \cite{YPR11}, and a similar version was used by \cite{D09}.

The current work presented in this paper uses our own sampler (described further in Section \ref{sec:implement}) based upon our interpretation of these ideas, implemented using our own blocking strategy. Our blocking strategy may or may not be original (we are unable to say given that the full blocking strategy adopted by \cite{YPR11} is not explicitly detailed), but we expect our approach to be based upon a sufficiently similar strategy such that the mixing issues that we
demonstrate would apply equally to other authors' implementations.

\subsection{An example model}
Equation \ref{eqn:dpmm2} is of course very general, indicating that sampling from the DPMM has wide scope across a variety of applications. However, it is perhaps equally instructive to consider a specific less abstract example, that can be used to highlight the issues raised in later sections.

\paragraph{Profile regression}
\label{subsec:profile}
Recent work has used the DPMM as an alternative to parametric regression, non-parametrically linking a response vector $\bY$ with covariate data $\bX$ by allocating observations to clusters. The clusters are determined by both the $\bX$ and $\bY$, allowing for implicit handling of potentially high dimensional interactions which would be very difficult to capture in traditional regression. The approach also allows for the possibility of additional ``fixed effects'' $\bW$ which have a global (i.e. non-cluster specific) effect on the response. The method is described in detail by \cite{MPJ10}, \cite{PMR11}, and \cite{MSM11}, who use the term \emph{profile regression} to refer to the approach. A similar model has independently been used by \cite{DHS08} and \cite{BD09}.

Using the notation introduced earlier in this Section, the data becomes $\bD = (\bY, \bX)$, and is modelled jointly as the product of a response model and and a covariate model resulting in the following likelihood:
\begin{equation}
\nonumber
 p(D_i|Z_i,\bTheta, \Lambda, W_i) = f_Y(Y_i|\Theta_{Z_i},\Lambda, W_i)f_X(X_i|\Theta_{Z_i}, \Lambda).
\end{equation}

\paragraph{Discrete covariates with binary response}
Consider the case where for each observation $i$, $X_i$ is a vector of $J$ locally independent discrete categorical random variables, where the number of categories for covariate $j=1,2,\ldots,J$ is $K_j$.
Then defining
\[
\Phi_c=(\phi_{c,1,1},\ldots,\phi_{c,1,K_1},\ldots,\phi_{c,J,1},\ldots,\phi_{c,J,K_J}),
\]
we specify the covariate model as:
\begin{equation}
\nonumber
   \mathbb{P}(X_i|Z_i,\Phi_{Z_i})=\prod_{j=1}^J\phi_{{Z_i},j,X_{i,j}}.
\end{equation}

Suppose also that $Y_i$ is a binary response, such that
\[
\mathrm{logit}\{\mathbb{P}(Y_i=1|\theta_{Z_i},\beta, W_i)\} = \theta_{Z_i} + \beta^{\mathrm{T}} W_i,
\]
for some vector of coefficients $\beta$.

This is simply an example of profile regression, with $\Theta_c = (\Phi_c,\theta_c)$ and $\Lambda = \beta$, such that
\[
 f_Y(Y_i|\Theta_{Z_i},\Lambda, W_i) = \mathbb{P}(Y_i|\theta_{Z_i},\beta, W_i),\;\;\textrm{and}
\]
\[
f_X(X_i|\Theta_{Z_i}, \Lambda) = \mathbb{P}(X_i|Z_i,\Phi_{Z_i}).
\]

We use this specific profile regression model to illustrate our results in this paper, both for the simulated dataset and the real-data example. Suitable prior distributions for making inference about such a model are discussed in \cite{MPJ10} and we adopt the same priors for the examples presented below. We note however that our conclusions and the behaviour we report typically hold more broadly across the range of models that we have tested.

\paragraph{Simulated datasets}
One of the key messages of our work is that DPMM samplers can perform well on simulated datasets but this does not necessarily carry through to real-data examples. We present in-depth results for a real-data example in Section \ref{sec:application}, but to highlight the contrasting performance two simple simulated dataset are also used. Our first simulated data is from a profile regression model with 10 discrete covariates and a binary response variable. The dataset has 1000 observations, partitioned at random into 5 groups in a balanced  manner.  The covariate and response distributions corresponding to each partition were selected to be well separated. The second simulated dataset is also from a profile regression model, but uses 10 discrete covariates, each with 5 categories, as well as 10 fixed effects and a Bernoulli outcome. However, in this case, the data is sampled by mixing over values of $\alpha$ from its Gamma prior, $\mathrm{Gamma}(9,0.5)$. An explicit description of the simulation methodology is provided in the Supplementary Material.

\section{Mixing of MCMC algorithms for the DPMM}
\label{sec:mixing}
Sampling from a DPMM is a non-trivial exercise, as evidenced by the number of different methods that have been introduced to address a wide array of issues. For Poly\`a Urn samplers, with mixture weights $\boldsymbol{\psi}$ integrated out, a primary limitation is that the conditional distribution of each cluster allocation variable depends explicitly upon all other cluster allocation variables. This means that the commonly used Gibbs samplers which typically update these variables one at a time suffer from poor mixing across partition space. Using Metropolis-within-Gibbs steps and bolder split-merge moves \citep{JN04} can improve results, but in high dimensional real-data applications, designing efficient moves of this type is far from straightforward.

The challenges associated with methods which sample from the FSBDPMM (most recently \citeauthor{YPR11}, \citeyear{YPR11} and \citeauthor{KGW11}, \citeyear{KGW11}) have been perhaps less well documented. This is partially because the innovative and ingenious methods that have facilitated such sampling have required significant attention in their own right, with the consequence that the methods are often illustrated only on relatively simple datasets.

The purpose of the remainder of this Section, and the main contribution of our work, is to use our practical experience to further understanding of the behaviour of this new type of samplers, with particular emphasis on some of the  challenges of sampling from the FSBDPMM for real data problems.

\subsection{Initial number of clusters}
A difficulty that persists even with the inclusion of the innovative techniques that allow MCMC sampling directly from the FSBDPMM is being able to effectively split clusters and thereby escape local modes. This is partially due to the intrinsic characteristics of partition spaces and the extremely high number of possible ways to split a cluster, even if it only has a small number (for example, 50 or more) subjects in it. Although sampling directly from the FSBDPMM (rather than integrating out the mixture weights) does improve mixing when updating the allocation variables, any Gibbs moves that update allocations and parameters individually (or even in blocks) struggle to explore partition space. On the other hand, constructing more ambitious Metropolis-Hastings moves that attempt to update a larger number of parameters simultaneously is also a very difficult task due to the difficulty in designing moves to areas of the model space with similar posterior support.

Rather than subtly ignoring the problem and reporting over confident inference when
analysing case studies, we suggest that, if used with caution,  a FSBDPMM sampler still provides a useful inferential tool, but that its limitations must be realised and acknowledged. For example, because of the difficulty that the sampler has in increasing the number of clusters for situations involving data with weak signal, it is important to initialise the algorithm with a number of clusters which is greater than the anticipated number of clusters that the algorithm will converge to. This necessarily involves an element of trial and error to determine what that number is, where multiple runs from different initialisations must be compared (for example using the ideas presented in Section \ref{sec:mpp}). This is demonstrated in Section \ref{sec:application}.

\subsection{Cluster ordering, $\alpha$ and label-switching}
\label{sec:label}
A secondary area where mixing of a full DPMM sampler requires specific attention is  the mixing of the algorithm over cluster orderings. In particular, whilst the likelihood of the DPMM is invariant to the order of cluster labels, the prior specification of the stick breaking construction is not. As detailed by \cite{PR08}, the definition of $\psi_c$ in terms of $V_c$, imposes the relation $\mathbb{E}[\psi_c]>\mathbb{E}[\psi_{c+1}]$ for all $c$. This weak identifiability, discussed in more detail by \cite{PIS08}, also manifests itself through the result $P(\psi_c>\psi_{c+1})>0.5$ for all $c$, a result that we prove in Appendix \ref{appendix:orderproof}.

The importance of whether the FSBDPMM algorithm mixes sufficiently across orderings depends partially upon the object of inference. Specifically, since $P(\psi_c>\psi_{c+1})$ depends upon the prior distribution of $\alpha$, if inference is to be simultaneously made about $\alpha$ (as is the scenario considered in this paper), it is very important that the algorithm exhibits good mixing with respect to $\alpha$. If this was not the case, the posterior marginal distribution for $\alpha$ would not be adequately sampled, and since $\alpha$ is directly related to the number of non-empty clusters (see \citeauthor{A74},\citeyear{A74} for details), poor mixing across ordering may further inhibit accurate inference being made about the number of non-empty clusters. This situation would be further exaggerated for more general stick breaking constructions (of the sort mentioned in the introduction). While it is possible to set a fixed value of $\alpha$, more generally we wish to allow $\alpha$ to be estimated.

To ensure adequate mixing across orderings, it is important to include label-switching moves, as observed by \cite{PR08}. Without such moves, the one-at-a-time updates of the allocations $Z_i$, mean that clusters rarely switch labels, and consequentially the ordering will be largely determined by the (perhaps random) initialisation of the sampler. For all choices of $\alpha$, the posterior modal ordering will be the one where the cluster with the largest number of individuals has label 1, that with the second largest has label 2 and so on. However, $\alpha$ affects the relative weight of other (non-modal) orderings, and a properly mixing sampler must explore these orderings according to their weights.

We adopt the label-switching moves suggested by \cite{PR08}, and details can be found therein. However, in our experience, while these moves may experience high acceptance rates early on in the life of the sampler, once a ``good'' (in terms of high posterior support) ordering is achieved, the acceptance rates drop abruptly (see Section 6, Figure \ref{fig:accRate1-3}) . This means that there is little further mixing in the ordering space. Our concern is that while these label-switching moves appear to encourage a move towards the modal ordering, once that ordering is attained, the sampler rarely seems to escape too far from this ordering.

Our solution is to introduce a third label-switching move that we describe here. In brief, the idea is to simultaneously propose an update of the new cluster weights so they are something like their expected value conditional upon the new allocations. Specifically, defining $Z^{\star} = \max_{1 \leq i \leq n} Z_i$ and $A = \{1,\ldots,Z^{\star}\}$ the move proceeds as follows: first choose a cluster $c$ randomly from $A\setminus\{Z^{\star}\}$. Propose new allocations
\begin{equation}
   \label{eqn:labelswitchZ}
   Z'_i=\begin{cases}
           c+1 & i:Z_i=c\\
           c & i:Z_i={c+1}\\
           Z_i & \textrm{otherwise}.
        \end{cases}
\end{equation}
and switch parameters associated to these clusters such that
\begin{equation}
   \label{eqn:labelswitchTheta}
   \Theta'_l=\begin{cases}
           \Theta_{c+1} & l=c\\
           \Theta_c & l=c+1\\
           \Theta_l & \textrm{otherwise}.
        \end{cases}
\end{equation}

Additionally, propose new weights $\psi'_c$ and $\psi'_{c+1}$ for components $c$ and $c+1$ such that
\begin{equation}
\label{eqn:labelswitchPsi}
   \psi'_l=\begin{cases}
            \psi_{c+1}\frac{\psi^+}{\Psi'}\frac{\mathbb{E}[\psi_c|\bZ',\alpha]}{\mathbb{E}[\psi_{c+1}|\bZ,\alpha]} & l=c\\
            \psi_c\frac{\psi^+}{\Psi'}\frac{\mathbb{E}[\psi_{c+1}|\bZ',\alpha]}{\mathbb{E}[\psi_c|\bZ,\alpha]} & l=c+1\\
            \psi_l & \textrm{otherwise,}
           \end{cases}
\textrm{and }
\end{equation}
where $\psi^+=\psi_c+\psi_{c+1}$ and
\[
   \Psi'=\psi_{c+1}\frac{\mathbb{E}[\psi_c|\bZ',\alpha]}{\mathbb{E}[\psi_{c+1}|\bZ,\alpha]}+\psi_c\frac{\mathbb{E}[\psi_{c+1}|\bZ',\alpha]}{\mathbb{E}[\psi_c|\bZ,\alpha]},
\]
by setting
\begin{equation}
\label{eqn:labelswitchV}
   V'_l=\begin{cases}
           \frac{\psi'_c}{\prod_{l<c}(1-V_l)} & l=c\\
           \frac{\psi'_{c+1}}{(1-V'_c)\prod_{l<c}(1-V_l)} & l=c+1\\
            V_l & \textrm{otherwise.}
        \end{cases}
\end{equation}
All other variables are left unchanged. Assuming that there are $n_c$ and $n_{c+1}$ individuals in clusters $c$ and $c+1$ respectively at the beginning of the update, the acceptance probability for this move is then given by $\min\{1,R\}$ where
\begin{eqnarray}
\label{eqn:labelswitchR}
 R & = & \left(\frac{\psi^+}{\psi_{c+1}R_1+\psi_{c}R_2}\right)^{n_c+n_{c+1}} R_1^{n_{c+1}} R_2^{n_c},\;\;\;\textrm{where }\\
\label{eqn:labelswitchR1}
R_1 & = & \frac{1+\alpha+n_{c+1}+\sum_{l>c+1}n_l}{\alpha+n_{c+1}+\sum_{l>c+1}n_l},\;\;\; \textrm{and}\\
\label{eqn:labelswitchR2}
R_2 & = & \frac{\alpha+n_c+\sum_{l>c+1}n_l}{1+\alpha+n_c+\sum_{l>c+1}n_l}.
\end{eqnarray}

More details can be found in Appendix \ref{appendix:labelswitch}.

\section{Monitoring convergence}
\label{sec:mpp}
Accepting that the challenge of convergence persists, it is clearly important that the user has diagnostic methods to assess whether convergence can be reasonably expected. Due to the nature of the model space, many traditional techniques cannot be used in this context. For our hierarchical model, as described in Equations \ref{eqn:stick} and \ref{eqn:dpmm2}, there are no parameters that can be used to  meaningfully demonstrate convergence of the algorithm. Specifically, parameters in the vector $\Lambda$ tend to converge very quickly, regardless of the underlying clustering, as they are not cluster specific and therefore are not a good indication of the overall convergence. On the other hand the cluster parameters $\bTheta_c$, cannot be tracked, as their number and interpretation changes from one iteration to the next (along with the additional complication that the labels of clusters may switch between iterations). While the concentration parameter $\alpha$ may appear to offer some information, using this approach can be deceiving, since a sampler that becomes stuck in a local mode in the clustering space will appear to have converged. Hence, monitoring the distribution of $\alpha$ across multiple runs initialised with different numbers of clusters is advisable, but in our experience finding a broad enough  spectrum of initialisations is not easy to determine in advance. Therefore, relying solely on $\alpha$ to monitor convergence might lead to misplaced confidence.

Based upon our experience with real datasets, we suggest that to better assess convergence, it is
also important to monitor the marginal partition posterior in each run, a calculation that we detail in the following section.

\subsection{Marginal partition posterior}
We define the marginal partition posterior as $p(\mathbf{Z}|\mathbf{D})$. This quantity represents the posterior distribution of the allocations given the data, having marginalised out all the other parameters.

In general computation of $p(\mathbf{Z}|\mathbf{D})$ is not possible in closed form, and requires certain assumptions and approximations. One such simplification is to fix the value of $\alpha$ in the calculation, rather than integrating over the distribution. Typically, we advise choosing one or several values of $\alpha$ to condition on, based on experimental runs on the dataset under study with $\alpha$ allowed to vary.

With the value of $\alpha$ fixed, whether or not $p(\mathbf{Z} |\mathbf{D})$ can be computed directly depends upon whether conjugate priors are adopted for all other parameter that must be integrated out. For the example of profile regression with logistic link introduced above this is typically not possible, as there is no natural conjugate for this response model. In such cases, integrating out such variables can be achieved using Laplace approximations. Using such an approximation appears to be sufficient for discerning differences between runs that perhaps indicate convergence problems. Details on the computations of $p(\mathbf{Z}|\mathbf{D})$ can be found in the Supplementary Material.

\begin{figure}
\centering
\includegraphics[width=8cm]{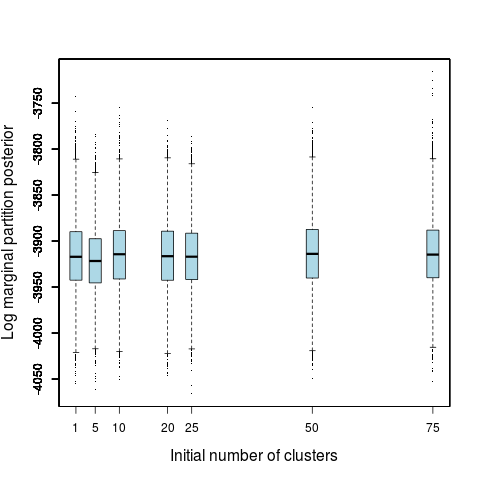}
\caption{Log marginal partition posterior for the first simulated dataset with different initial number of clusters.\label{fig:margModelPosteriorSim}}
\end{figure}

Figure \ref{fig:margModelPosteriorSim} demonstrates that the strong signal in our first simulated dataset means that the sampler converges regardless of the initial number of clusters. In contrast, Section \ref{sec:application} (Figure \ref{fig:alphaByInitClus3})  demonstrates that for our real dataset convergence is not always achieved.

Computing the marginal partition posterior for each run of the MCMC and comparing between runs has proven to be a very effective tool for our real examples, particularly to identify runs that were significantly different from others, perhaps due to convergence issues.

Whereas comparing the marginal distribution of a parameter such as $\alpha$ between MCMC runs might help diagnose non-convergence if used with a wide range of initialisations, it gives no indication of which run has explored the regions of higher posterior probability. On the other hand, comparing the marginal partition posterior between two differing runs immediately indicates which run explored the higher posterior probability regions. This means that even if we are not able to make fully Bayesian inference about the parameters, we are able to draw some conclusions about those parameters which are more likely.

\section{Our implementation of a DPMM sampler}
\label{sec:implement}
To demonstrate the behaviour discussed within this paper, we have used our own implementation of a Gibbs sampler (with Metropolis-within-Gibbs steps) for the FSBDPMM. The core of the sampler is implemented as efficient \texttt{C++} code, interfaced through the \texttt{PReMiuM} \texttt{R} package \citep{LHR13}.

The sampler was originally written specifically for analysis of profile regression problems (as presented in Section \ref{subsec:profile}) across a variety of applications. For such models, the package includes Bernoulli, Binomial, Poisson, Normal and categorical response models, as well as Normal and discrete covariates. It is also possible to run the sampler with no response model, allowing the consideration of more traditional mixture models. Additionally, the sampler implements a type of variable selection, allowing inference to be made in the case of data where the clustering might be determined with reference to only a subset of covariates This type of problem is discussed in detail by \cite{PMH12}.

Extensive details of the algorithm can be found in \citep{LHR13}, including the blocking strategy that is integral for allowing sampling from the FSBDPMM. We note some brief details that are relevant to the current work below.

\subsection{Post processing}
\label{sec:postprocessing}
\paragraph{An optimal partition} Given a sample of partitions from the posterior distribution of a Bayesian cluster model (for example from a DPMM sampler where the sample is the output of an MCMC algorithm) it is often desirable to summarise the sample as a single representative clustering estimate. The benefits of having a single estimate of the partition often sufficiently outweigh the fact that the uncertainty of the clustering is lost by such a point estimate, although it should always be communicated that this uncertainty may be considerable.

One benefit of using an optimal partition is that questions of how to account for unambiguous labelling of clusters between MCMC sweeps can be avoided, which would not be the case if we wished to provide certain kinds of distributional summary of the partition space. We emphasise that the term label-switching is often used in this context to refer to the complicating impact on inference of not having ways of ``tracking'' clusters between iterations. This is in contrast to  the deliberate label-switching moves as introduced in Section \ref{sec:label} which use label-switching as a technique to better explore partition space and avoid undue influence of the ordering. Note that our inferential methods (e.g. determining an optimal partition or the predictive method described in the following section) are not affected by label-switching.

There are many different ways to determine a point estimate of the partition, for example something as simple as the maximum a posteriori (MAP) estimate (the partition in the sample with the highest value of the marginal partition posterior). We prefer methods based on the the construction (as a post-processing step) of a posterior similarity matrix, a matrix containing the posterior probabilities (estimated empirically from the MCMC run) that the observations $i$ and $j$ are in the same cluster. The idea is then to find a partition which maximises the sum of the pairwise similarities. We find that methods based on the posterior similarity matrix are less susceptible to Monte Carlo error, especially when the optimal partition is not constrained to be in sample, but might be obtained using additional clustering methods, such as partitioning around medoids, that take advantage of the whole MCMC output. Note that once a representative partition is chosen, full uncertainty about its characteristic features can be recovered from postprocessing of the full MCMC output. See \citep{MPJ10} for a full discussion.

\paragraph{Making predictions}
While an optimal partition can be very helpful in some cases (particularly when it is the clustering itself that is the primary object of inference) difficulties are faced in understanding or conveying the uncertainty of the partitioning. Due to the complexity and sheer size of the model space, the optimal partitions tend to differ between runs of the MCMC, and it is not an easy task to assess whether convergence has been achieved based on this approach alone.

A common target of inference is not necessarily the partition itself, but how the estimated parameters might allow us to make predictions for future observations. For example we might want to group new observations with existing observations, or, in the case of profile regression, make a prediction about the response if only the covariates of a new observation had been observed. One way to do this is to use posterior predictions, where posterior predictive distributions for quantities of interest can be derived from the whole MCMC run, taking the uncertainty over clustering into account.

Depending on the quantity of interest, the posterior predictive distribution can often be relatively robust even across runs with noticeably different optimal partitions. While this may not help us to determine if the algorithm has sufficiently explored the partition-space, if the purpose of the inference is to make predictions, this robustness can be reassuring. Moreover, by allowing predicted values to be computed based on probabilistic allocations (i.e. using a Rao-Blackwellised estimate of predictions) results can be further desensitised to a single best partition.

\section{Investigation of the algorithm's properties in a large data application}
\label{sec:application}
In this section, we report the results of using our FSBDPMM sampler in a profile regression application with discrete covariates and a binary  response, applied to a real epidemiological dataset with 2,639 subjects.

The analysis of real data presents a number of challenges: it requires care in ensuring convergence, as the signal is not as strong as in a simulation study. However, these are challenges that might be encountered more widely by users wishing to apply the methods to real data, and by presenting an example it allows us to highlight and discuss the issues that arise.

\subsection{The data}
Our dataset is a subset taken from an epidemiological case-control study, the analysis of which has provided the motivation of most of the work presented in this paper (see \citeauthor{HLA13}, \citeyear{HLA13}). In the illustrative example we have 2,639 subjects, and use 6 discrete covariates each with 5 categories, and 13 fixed effects. The response is binary and we use the model specifications detailed in Section 2.2 to analyse this data set. The low signal contained in the data poses issues with convergence of the MCMC, as we illustrate below.

Our results are based upon running multiple chains each for 100,000 iterations after a burn-in sample of 50,000 iterations. In some cases, behaviour within this burn-in period is illustrated.

\subsection{Results}
\label{sec:results}
\begin{figure}
\centering
\includegraphics[width=8cm]{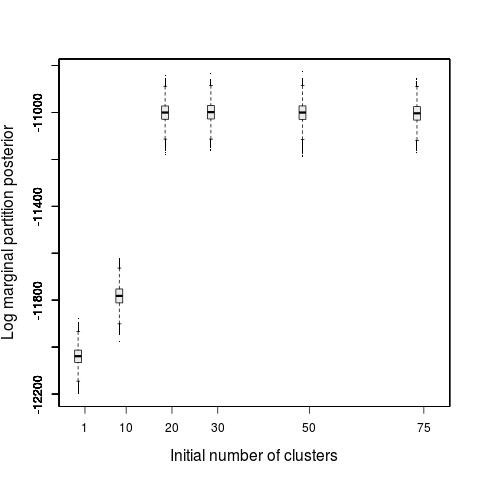}
\caption{Log marginal partition posterior for the real epidemiological dataset with different initial number of clusters.\label{fig:margModelPosterior}}
\end{figure}

\paragraph{Marginal partition posterior and number of clusters}
As discussed in Section \ref{sec:mixing} we run multiple MCMC runs, starting each with very different numbers of initial clusters. For this dataset, initialising the sampler with fewer than 20 clusters results in marginal partition posterior distributions that are significantly different between runs. This is illustrated in Figure \ref{fig:margModelPosterior}, where initialisations with small number of clusters result in much lower marginal partition posterior values than can be achieved with a higher initial number of clusters. It is apparent that there is a cut-off at 20 clusters, where increasing the number of initial clusters further does not result in an increase in the marginal partition posterior, suggesting that with 20 clusters or more the sampler is able to visit areas of the model space with the highest posterior support.

\begin{figure}
\centering
\includegraphics[width=8cm]{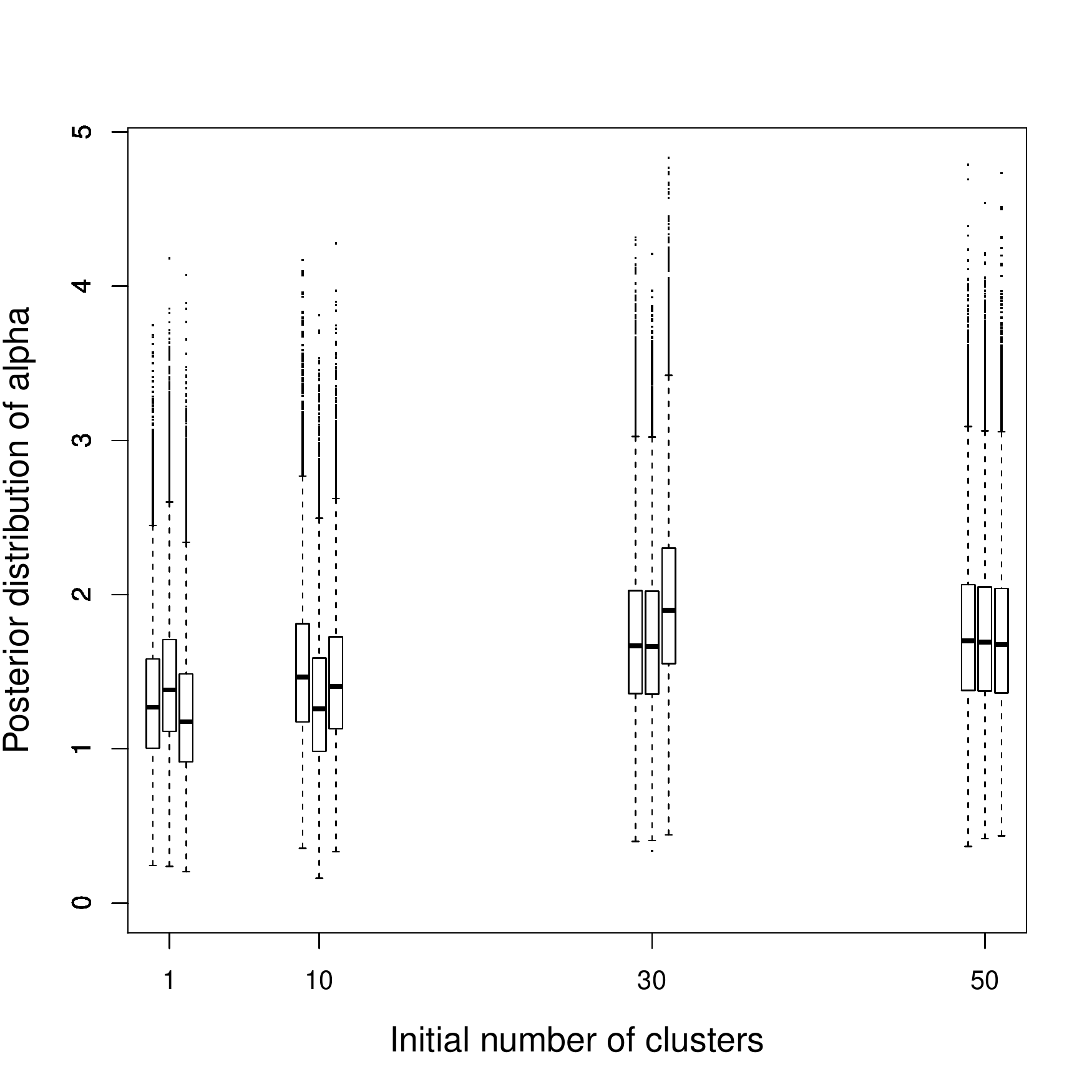}
\caption{Posterior distribution of $\alpha$ for different number of initial clusters with three repetitions per initialisation: boxplots for the distribution for 50,000 sweeps after a burn-in of 50,000 samples.\label{fig:alphaByInitClus3}}
\end{figure}

\paragraph{Posterior distribution of $\alpha$}
Figure \ref{fig:alphaByInitClus3} shows the boxplot of the posterior distribution of $\alpha$ as a function of the initial number of clusters. For each different initial number of clusters, three different runs with random initialisations of other parameters were performed. We can see that the posterior distribution of $\alpha$ only stabilises when the initial number of clusters is high, around 50 in our case. Thus, we would recommend carrying out such checks as part of the investigation of convergence strategy. Note that while it is advisable to start with a large number of initial clusters, starting with many more clusters than necessary can result in a larger number of iterations required for convergence.

\begin{figure}
\centering
\includegraphics[width=8cm]{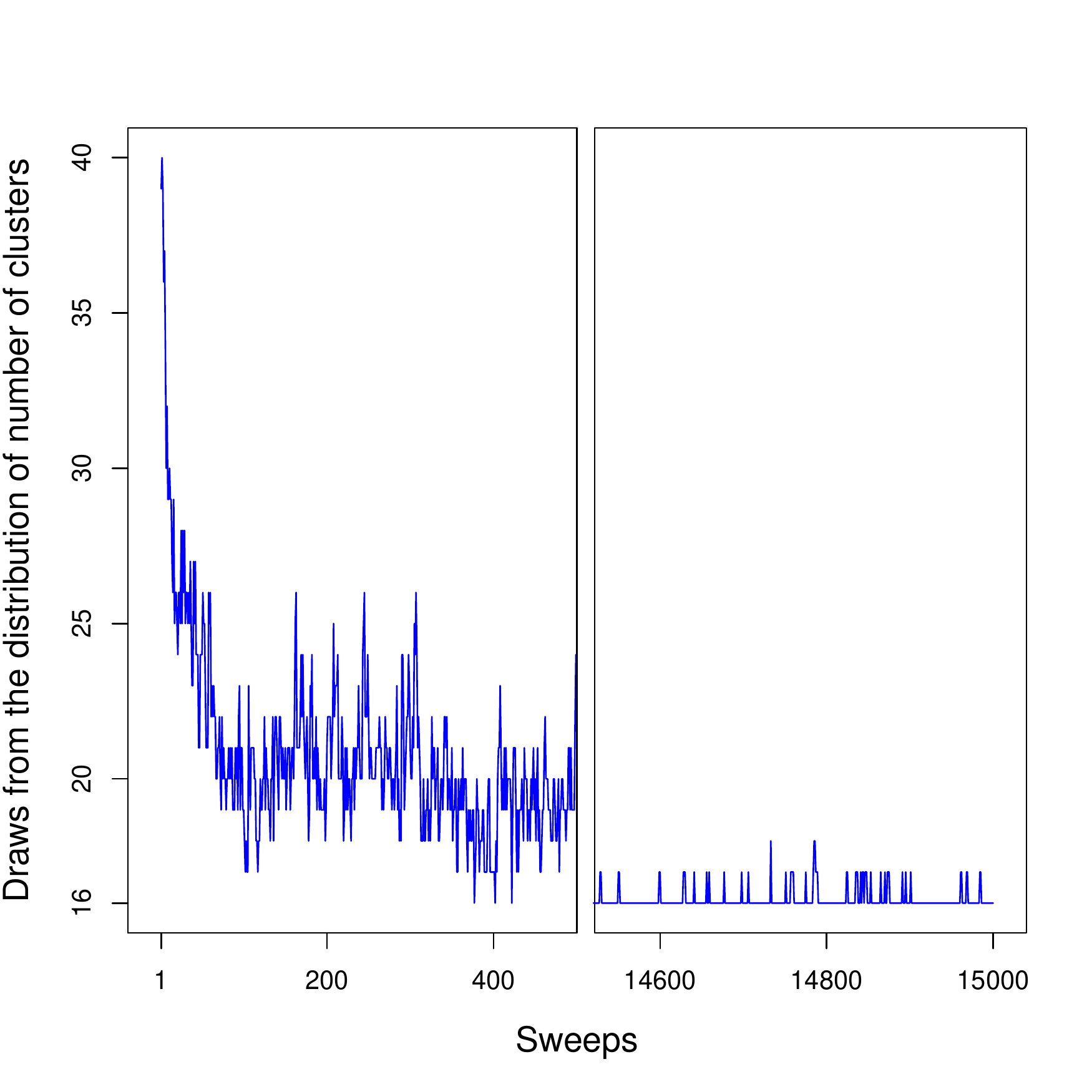}
\caption{The trace of the posterior of the number of clusters for the first 500 iterations and after 15,000 iterations of the MCMC sampler.\label{fig:newFig5-6}}
\end{figure}

\paragraph{Posterior distribution of the number of clusters}
The need to initialise the sampler with a sufficiently high number of clusters is also supported by
looking at the posterior distribution of the number of clusters. Figure \ref{fig:newFig5-6} contrasts the behaviour of the sampler between the first 500 iterations of the burn in period  and 500 iterations after the first 15,000, for a run with 31 initial clusters.

In the initial iterations, the space is explored by modifying and merging clusters, with the number of clusters changing frequently, in a general downward trend. On the other hand, once the MCMC has converged to the model space around a mode, the algorithm attempts to split clusters regularly, but the number of changes in the number of clusters are few, and increases in the number of clusters are almost immediately reversed in the following iteration.

\begin{figure}
\centering
\includegraphics[width=8cm]{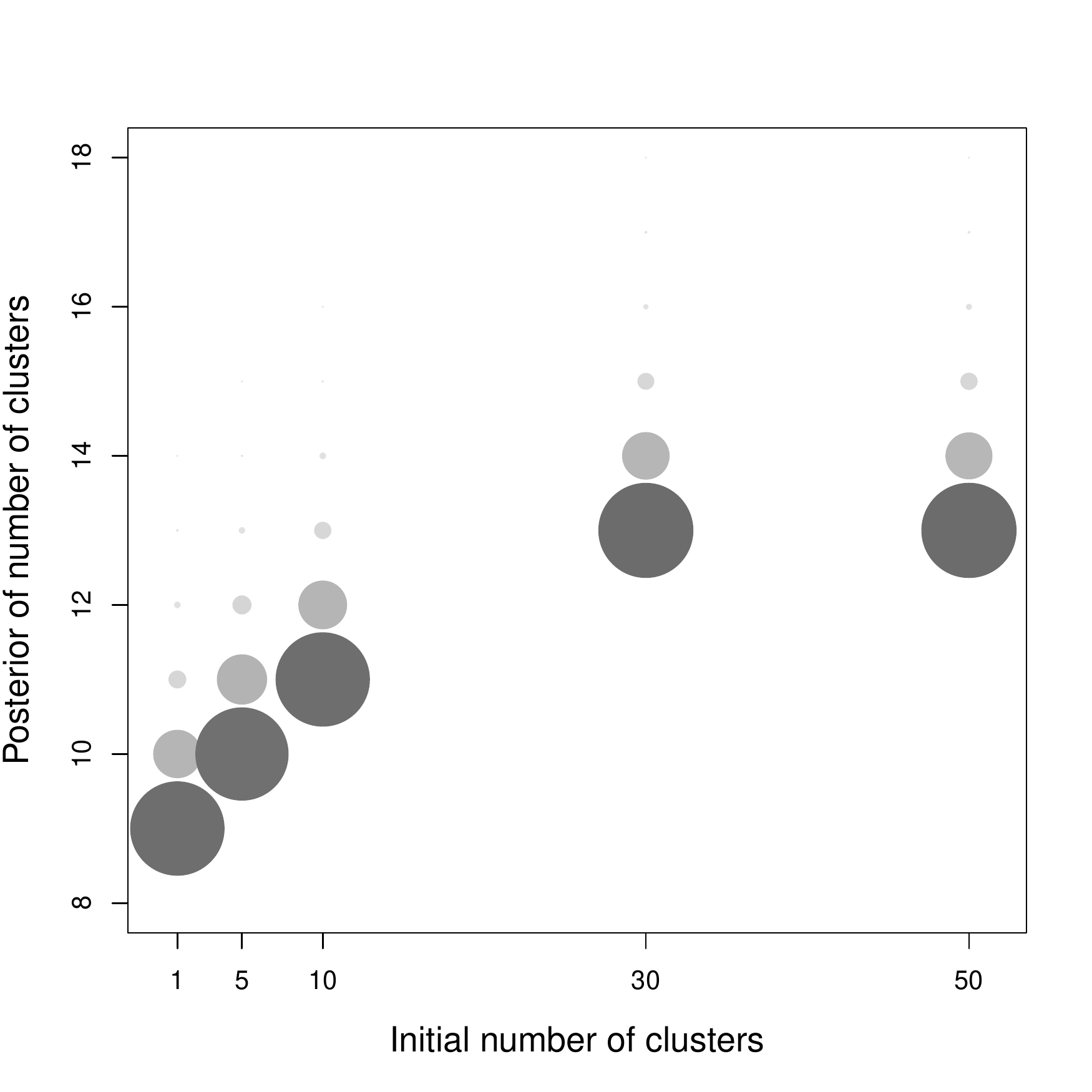}
\caption{The posterior distribution of the number of clusters for 50,000 sweeps after a burn-in of 50,000 iterations.\label{fig:NEWnClusters3c}}
\end{figure}

The posterior distributions for the number of clusters is shown in Figure \ref{fig:NEWnClusters3c} for runs with different initial numbers of clusters. Five chains have been ran, initialised with 1, 5, 10, 30 and 50 clusters respectively. The size and shading of each circle in Figure \ref{fig:NEWnClusters3c} represents the posterior frequency of the number of clusters for each of the chains. As can be seen from this figure, with 30 or more initial clusters the sampler has converged to a common area of posterior support, but with fewer than this the sampler might not visit this region of the model space, despite it having increased posterior support. Taken together, the plots in Figures\ref{fig:margModelPosterior}, \ref{fig:alphaByInitClus3} and \ref{fig:NEWnClusters3c} provide concurring evidence that for our real data case, starting with 50 or more clusters leads to reproducible conclusions.

\begin{figure}
\centering
\includegraphics[width=8cm]{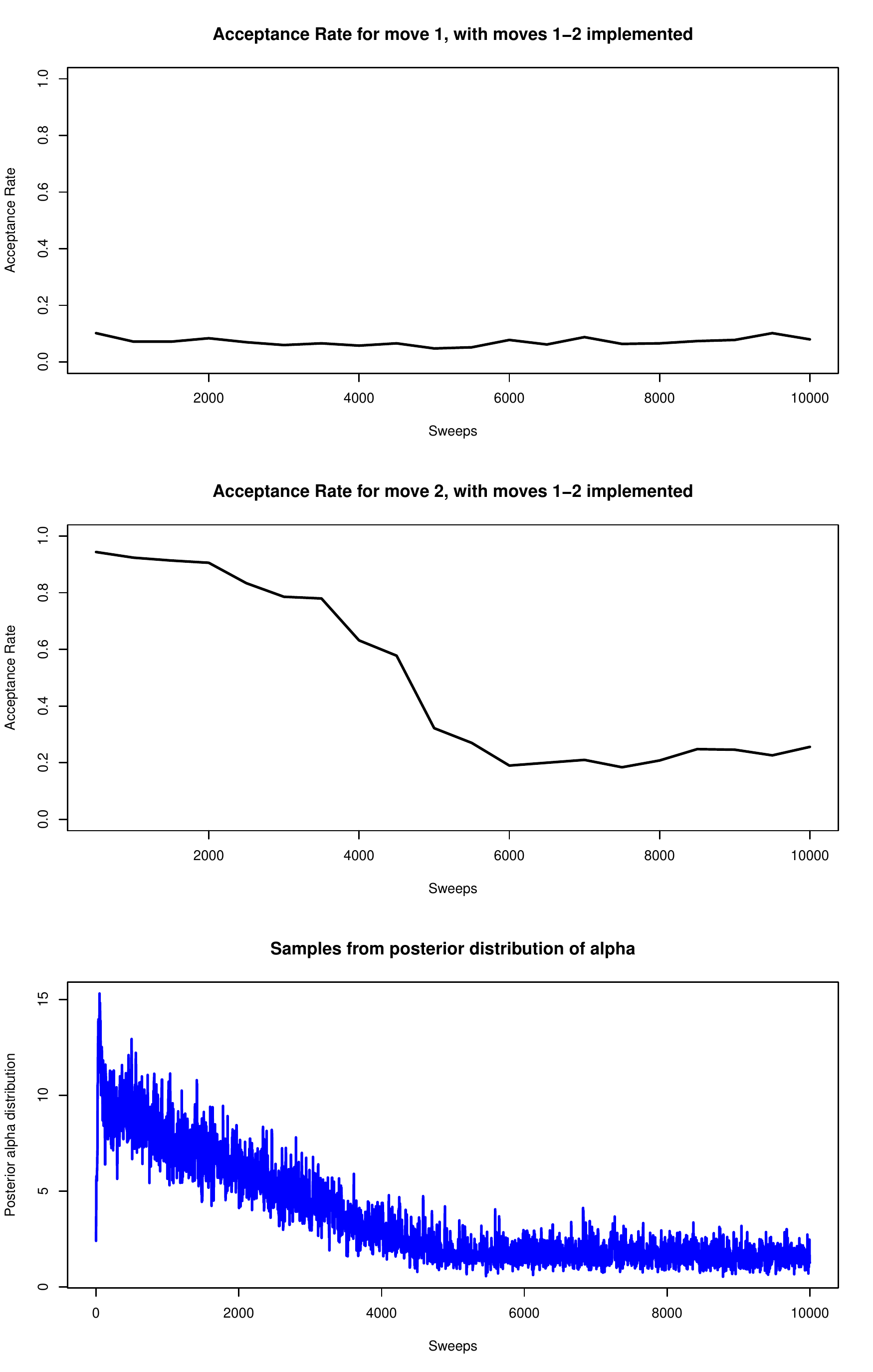}
\caption{\label{fig:accRate}Acceptance rate for intervals of 500 sweeps for the two label-switching moves proposed by \cite{PR08} and comparison with samples from the posterior distribution of $\alpha$ (bottom).}
\end{figure}

\paragraph{Label-switching moves}
This example also demonstrates the need for the new label-switching move discussed in Section \ref{sec:label} to ensure good mixing. Figure \ref{fig:accRate} demonstrates the decrease in acceptance rate that is evidenced for the label-switching moves, if only the moves that \cite{PR08} propose are included. For the first of the moves that \cite{PR08} propose, where the labels of two randomly selected clusters are exchanged, we observed acceptance rates below 10\% for any sample of 500 sweeps. For the second of the moves, where the labels of two neighbouring clusters are swapped, along with the corresponding $V_c$, $V_{c+1}$ the acceptance rate drops considerably after initially being very high.    This decrease can be explained by the observation (made by the original authors) that the second move type is always accepted if one of the clusters is empty, which can happen often in initial cluster orderings with low posterior support. Note that $\alpha$ stabilises after 5,000 iterations for the example shown. If only the first of the two moves is implemented, $\alpha$ moves extremely slowly
(more than 50,000 iterations are not enough to have a stable trace; not shown) while if only the second of the two moves is implemented, for this example, 17,000 iterations are necessary for $\alpha$ to stabilise (not shown).

\begin{figure}
\centering
\includegraphics[width=8cm]{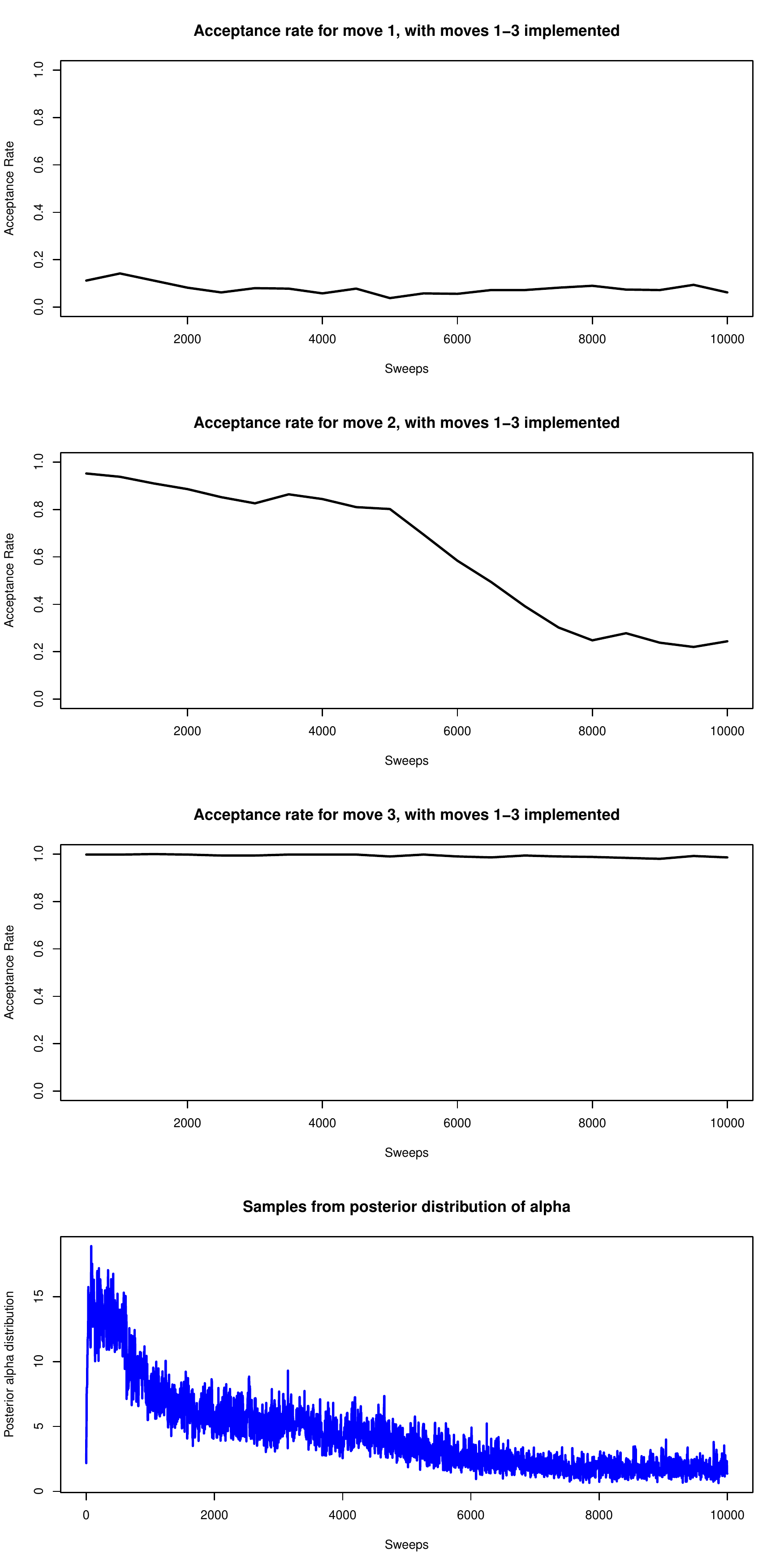}
\caption{Acceptance rates with the new label-switching move.\label{fig:accRate1-3}}
\end{figure}

\begin{figure}
\centering
\includegraphics[width=8cm]{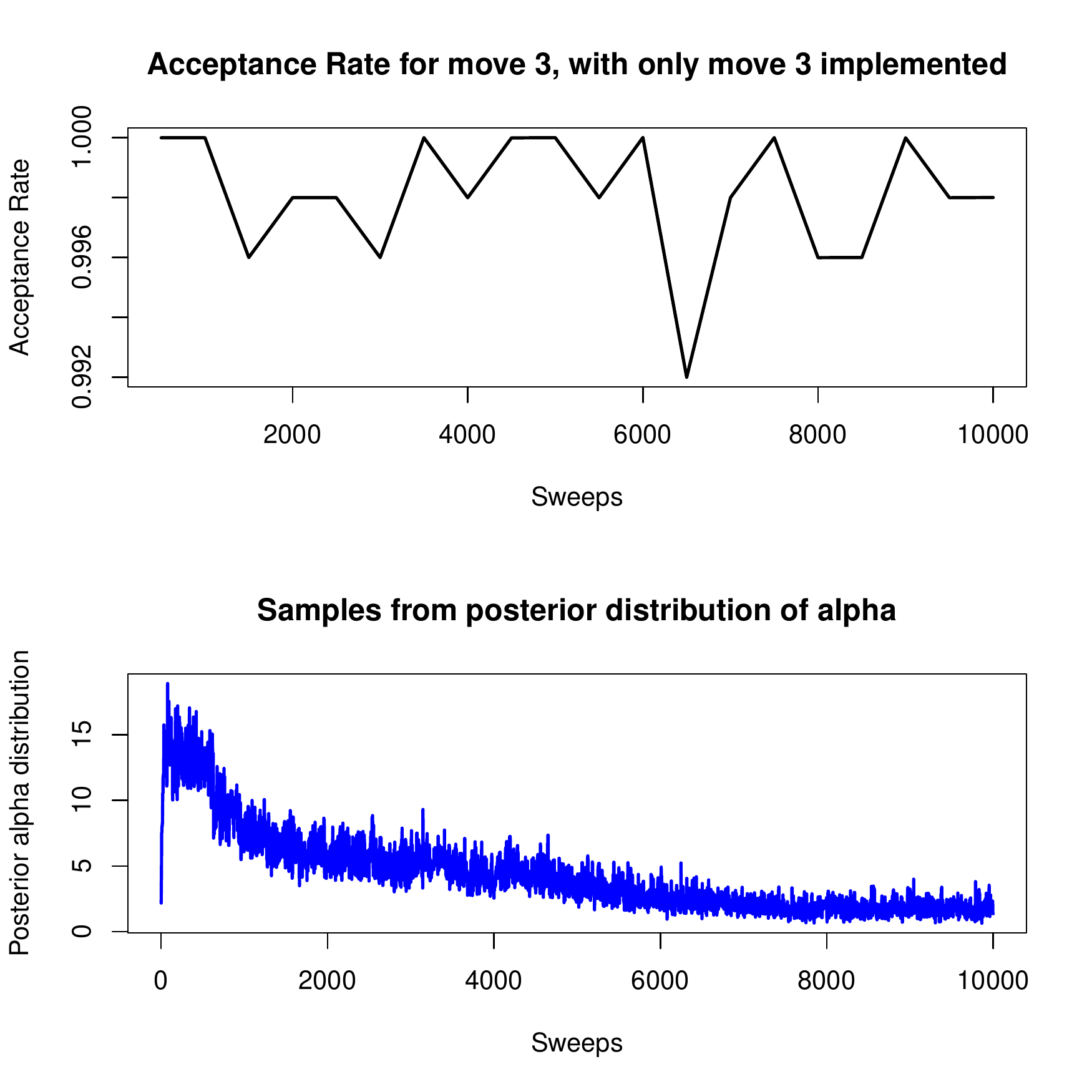}
\caption{Acceptance rates for the new label-switching move.\label{fig:accRate3}}
\end{figure}

Comparing Figure \ref{fig:accRate1-3} to Figure \ref{fig:accRate}, we can see that the new label-switching move suffers from no drop off in acceptance at any point throughout the run. Figure \ref{fig:accRate3} shows the acceptance rate for our new label-switching move, when the other two switching label is not implemented. While the performance is worse than using all three moves, it is the most effective single label-switching move (see Section \ref{sec:label}).

\begin{figure}
\centering
\includegraphics[width=8cm]{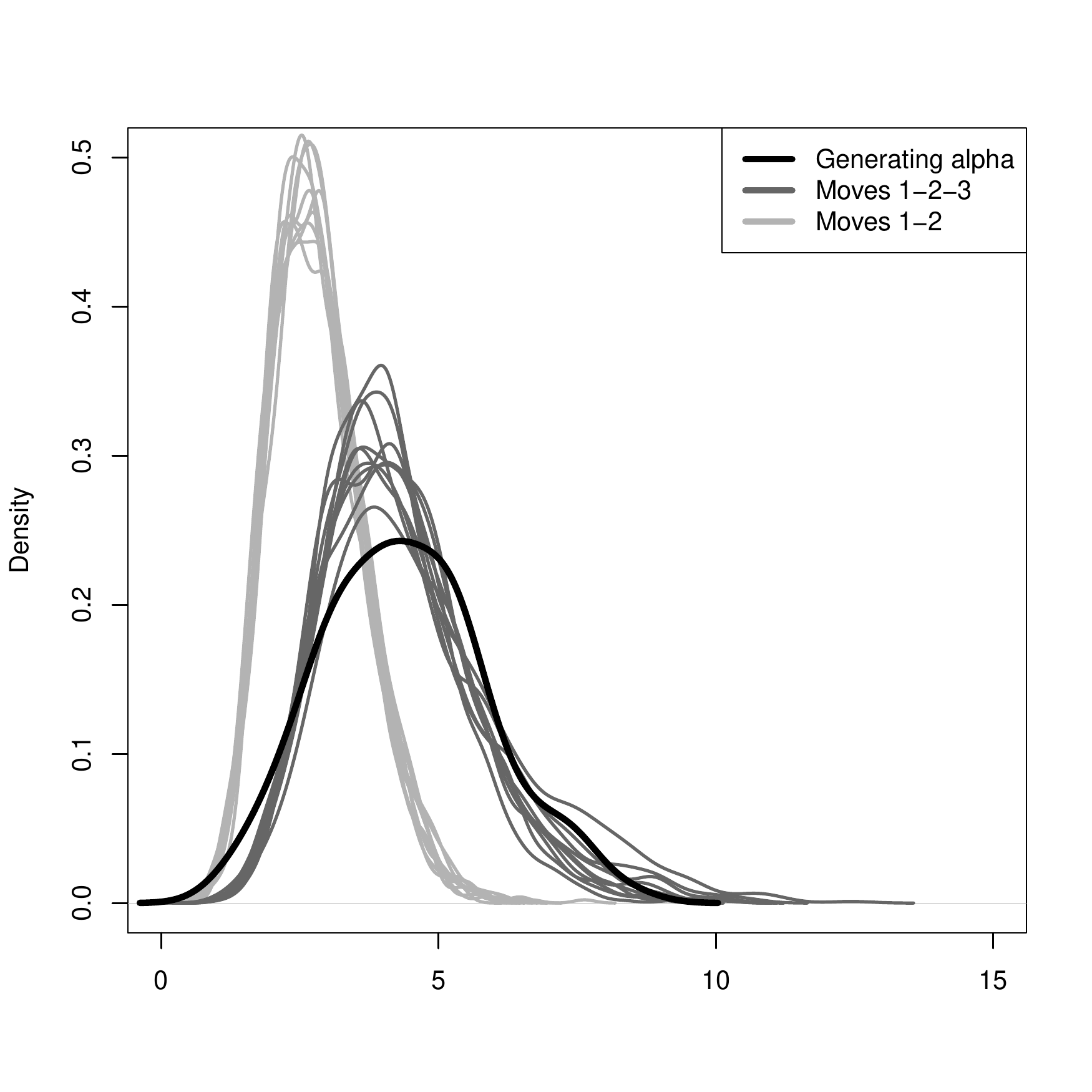}
\caption{Recovered posterior density of $\alpha$ from multiple MCMC runs with and without the new label-switching move compared with generating density of $\alpha$ for the second simulated dataset. \label{fig:simRandomAlphaDensity}}
\end{figure}

To further assess how the new label-switching move affects mixing and the ability to recover
 the posterior distribution of $\alpha$, we used our second simulated dataset. Starting with 100 clusters, we performed 10 runs of the sampler using only moves 1 and 2 for label-switching, and 10 runs adding in our third label-switching move. In each case we ran the chain for 100,000 iterations after a burn-in sample of 100,000 iterations. Figure \ref{fig:simRandomAlphaDensity} shows the performance of the sampler in retrieving the distribution of $\alpha$ that was used to simulate the data with and without using our new label-switching move. It is clear that this distribution is not well recovered when using exclusively moves 1 and 2, while the addition of our third label-switching move is clearly beneficial.

\section{Conclusions}
\label{sec:conclusions}
By demonstrating some of the challenges that occur when sampling from the DPMM, we hope to have raised awareness that continued research into the DPMM sampling methodology is required. Our implementation of a FSBDPMM sampler, synthesises many of the most recent and innovative techniques introduced by other authors, such as parameter blocking, slice sampling, and label-switching. However, due to the complex model space that is inherent with the FSBDPMM, many issues persist.

In previous work by other authors, considerable progress has been made evolving the samplers through innovative strategies and approaches. Nonetheless, discussion of many of the residual difficulties is avoided through demonstrating the methods only on simulated data, or for datasets with strong signal. In practice however, with real datasets, the user does not have the option of simply avoiding these issues, as illustrated by our analysis of the mixing performance of an epidemiological data set with low signal to noise ratio.

In this paper we have attempted to highlight the difficulties that a user may face in practice. We have added a new feature in the form of an additional label-switching move to build upon this previous research and further alleviate some of the challenges that are involved when trying to sample such a complex posterior space. We have also provided practical guidelines based on our experience, on how to make useful inference in the face of these limitations.

As a consequence of discussing these challenges explicitly, we hope that our work will motivate further developments in this area to take additional steps to improve sampler efficiency. The challenge of designing MCMC moves that are able to escape local well-separated modes is considerable, but equally, so is the imagination and innovation of many researchers developing new MCMC sampling methodologies. Encouragingly research continues, and drawing on alternative techniques which might be better designed for multi-modality, such as sequential Monte Carlo (see for example \citeauthor{UGC10}, \citeyear{UGC10}) may yield further improvements.

In the meantime,  practitioners may benefit from observing the difficulties we have presented here, allowing them to recognise and communicate potential limitations of their analyses.

\section*{Acknowledgements}
David I. Hastie acknowledges support from the INSERM grant (P27664). Silvia Liverani acknowledges support from the Leverhulme Trust (ECF-2011-576). Sylvia Richardson acknowledges support from MRC grant G1002319. We are grateful for helpful discussions with Sara K. Wade and Lamiae Azizi, and to Isabelle St\"{u}cker for providing the epidemiological data. We would like to the thank the editor and reviewers for their helpful comments that have allowed us to improve this paper.

\begin{appendix}
\section{Appendices}

\subsection{}
\label{appendix:orderproof}
We provide the following proposition concerning the relationship between the ordering and $\alpha$.

\begin{proposition}
Suppose that we have a model with posterior as given in Equation \ref{eqn:stick}. Then $\mathbb{P}(\psi_c>\psi_{c+1}|\alpha)$ is a function of $\alpha$, and furthermore $\mathbb{P}(\psi_c>\psi_{c+1})>0.5$.
\begin{proof}
If $\psi_c>\psi_{c+1}$ then $V_c > V_{c+1}(1-V_c)$, which implies $V_{c+1}<V_c/(1-V_c)$. Thus
\begin{eqnarray*}
   \mathbb{P}(&\psi_c&>\psi_{c+1}|\alpha) = \mathbb{P}(V_{c+1}<V_c/(1-V_c)|\alpha)\\
& = & \int_{0}^{0.5}\int_{0}^{V_1/(1-V_1)}\alpha^2(1-V_1)^{\alpha-1}(1-V_2)^{\alpha-1}\mathrm{d}V_2\mathrm{d}V_1\\
& &\;\;\;\;\;\;\; +\;\;\; \int_{0.5}^{1}\int_{0}^{1}\alpha^2(1-V_1)^{\alpha-1}(1-V_2)^{\alpha-1}\mathrm{d}V_2\mathrm{d}V_1\\
& = & \int_{0}^{0.5}\left[\alpha(1-V_1)^{\alpha-1}-\alpha(1-V_1)^{\alpha-1}\left(\frac{1-2V_1}{1-V_1}\right)^{\alpha}\right]\mathrm{d}V_1 \\ &&\;\;\;+\int_{0.5}^1 \alpha(1-V_1)^{\alpha-1}\mathrm{d}V_1\\
& = & \int_{0}^{1}\alpha(1-V_1)^{\alpha-1}\mathrm{d}V_1 \\ && \;\;\;- \int_{0}^{0.5}\alpha\frac{(1-2V_1)^{\alpha}}{1-V_1}\mathrm{d}V_1\\
& = & 1- \int_{0}^{0.5}\alpha\frac{(1-2V_1)^{\alpha}}{1-V_1}\mathrm{d}V_1.
\end{eqnarray*}
Now since, $(1-2V_1)^{\alpha}/(1-V_1) < (1-2V_1)^{\alpha-1}$
\[
   \alpha\int_{0}^{0.5}\frac{(1-2V_1)^{\alpha}}{1-V_1}\mathrm{d}V_1 < \alpha\int_{0}^{0.5}(1-2V_1)^{\alpha-1}\mathrm{d}V_1 =0.5.
\]
So $\mathbb{P}(\psi_c>\psi_{c+1}|\alpha)>0.5$ for all $\alpha$. Finally,
\begin{eqnarray*}
   \mathbb{P}(\psi_c>\psi_{c+1}) &=& \int\mathbb{P}(\psi_c>\psi_{c+1}|\alpha)p(\alpha)\mathrm{d}\alpha\\& >&\int 0.5 p(\alpha)\mathrm{d}\alpha = 0.5.
\end{eqnarray*}
\end{proof}
\end{proposition}

\subsection{}
\label{appendix:labelswitch}
\begin{proposition}
Consider the label-switching move defined in Equations \ref{eqn:labelswitchZ} to \ref{eqn:labelswitchV} in Section \ref{sec:label}. Then:
\begin{enumerate}
 \item[(i)] $(\psi^+)':=\psi'_c+\psi'_{c+1}=\psi_c+\psi_{c+1}=\psi^+$;
 \item[(ii)] $(1-V'_c)(1-V'_{c+1})=(1-V_c)(1-V_{c+1})$;
 \item[(iii)] The proposal mechanism is its own reverse;
 \item[(iv)] \[\frac{\mathbb{E}(\psi_c|\bZ',\alpha)}{\mathbb{E}(\psi_{c+1}|\bZ,\alpha)} = \frac{1+\alpha+n_{c+1}+\sum_{l>c+1}n_l}{\alpha+n_{c+1}+\sum_{l>c+1}n_l}\;\;\;\textrm{and}\]
	     \[\frac{\mathbb{E}(\psi_{c+1}|\bZ',\alpha)}{\mathbb{E}(\psi_c|\bZ,\alpha)} = \frac{\alpha+n_c+\sum_{l>c+1}n_l}{1+\alpha+n_c+\sum_{l>c+1}n_l};\;\;\;\textrm{and}\]
 \item[(v)] the acceptance probability for this move is given by $\min\{1,R\}$, where the acceptance ratio $R$ is given in Equation \ref{eqn:labelswitchR}.

\end{enumerate}

\begin{proof}
\begin{enumerate}
 \item[(i)] By definition
\begin{eqnarray*}
(\psi^+)'&:=&\psi'_c+\psi'_{c+1}\\
&=&\frac{\psi^+}{\Psi'}\left(\psi_{c+1}\frac{\mathbb{E}[\psi_c|\bZ',\alpha]}{\mathbb{E}[\psi_{c+1}|\bZ,\alpha]}+\psi_c\frac{\mathbb{E}[\psi_{c+1}|\bZ',\alpha]}{\mathbb{E}[\psi_c|\bZ,\alpha]}\right)\\&=&\frac{\psi^+}{\Psi'}\Psi'=\psi^+;
\end{eqnarray*}
 \item[(ii)] From (i),
\begin{eqnarray*}
\psi'_c+\psi'_{c+1} & = & \psi_c+\psi_{c+1}
\end{eqnarray*}
implies
\begin{eqnarray*}
\left[V'_c+V'_{c+1}(1-V'_c)\right]&& \prod_{l<c}(1-V'_l) \\&=& \left[V_c+V_{c+1}(1-V_c)\right] \prod_{l<c}(1-V_l).
\end{eqnarray*}
By Equation \ref{eqn:labelswitchV}, $V'_l=V_l$ for all $l<c$,
\begin{eqnarray*}
\Rightarrow\;\;\;V'_c+V'_{c+1}(1-V'_c) & = & V_c+V_{c+1}(1-V_c)\\
\Rightarrow \;\;\;(1-V'_c)(1-V'_{c+1}) & = & (1-V_c)(1-V_{c+1}).
\end{eqnarray*}
The importance of this result is that it provides confirmation that our proposed $\psi'$ in Equation \ref{eqn:labelswitchPsi} can be achieved with the $V$ defined in Equation \ref{eqn:labelswitchV}. In particular, with this choice of $V'$, the only weights that are changed are those associated with components $c$ and $c+1$, as desired.
 \item[(iii)] Suppose that the Markov chain is currently in the proposed state defined in Equations \ref{eqn:labelswitchZ} to \ref{eqn:labelswitchV} i.e.
$(\bV',\bTheta',\bZ',\bU,\alpha,\Lambda)$. We show that applying the proposal mechanism to this state, for component $c$ and $c+1$, the proposed new state is the original state \[(\bV'',\bTheta'',\bZ'',\bU,\alpha,\Lambda)=(\bV,\bTheta,\bZ,\bU,\alpha,\Lambda.)\]

The parameters $\bU$, $\alpha$ and $\Lambda$ are unchanged by design of the proposal mechanism. Also, by design, the allocations $\bZ$ and cluster parameters $\bTheta$ are simply swapped for the selected components, so trivially $\bZ''=\bZ$ and $\bTheta''=\bTheta$. Since $V''_l$ is unchanged for $l\not\in\{c,c+1\}$, it remains only to show $V''_c=V_c$ and $V''_{c+1}=V_{c+1}$, or equivalently $\psi''_c=\psi_c$ and $\psi''_{c+1}=\psi_{c+1}$. To confirm,
\begin{eqnarray}
\nonumber
 \psi''_c & = & \psi'_{c+1}\frac{(\psi^+)'}{\Psi''}\frac{\mathbb{E}[\psi_c|\bZ'']}{\mathbb{E}[\psi_{c+1}|\bZ',\alpha]}\\
\nonumber
 & = & \psi_c\frac{\psi^+}{\Psi''}\frac{\psi^+}{\Psi'}\frac{\mathbb{E}[\psi_{c+1}|\bZ',\alpha]}{\mathbb{E}[\psi_c|\bZ,\alpha]}\frac{\mathbb{E}[\psi_c|\bZ'',\alpha]}{\mathbb{E}[\psi_{c+1}|\bZ',\alpha]}\\&&(\textrm{by (i) and Equation \ref{eqn:labelswitchPsi}})\\
\label{eqn:psidashdash}
& = & \psi_c\frac{\left(\psi^+\right)^2}{\Psi''\Psi'}\;\;\;\textrm{since }\bZ''=\bZ.
\end{eqnarray}
However,
\begin{eqnarray*}
 \Psi'' & = & \psi'_{c+1}\frac{\mathbb{E}[\psi_c|\bZ'']}{\mathbb{E}[\psi_{c+1}|\bZ',\alpha]}+\psi'_c \frac{\mathbb{E}[\psi_{c+1}|\bZ'',\alpha]}{\mathbb{E}[\psi_c|\bZ',\alpha]}\\
        & = & \frac{\psi^+}{\Psi'}(\psi_c+\psi_{c+1}) \\&& (\textrm{from Equation \ref{eqn:labelswitchPsi} and since }\bZ''=\bZ )\\
        & = & \frac{\left(\psi^+\right)^2}{\Psi'}.	
\end{eqnarray*}
Substituting this into Equation \ref{eqn:psidashdash} we get $\psi''_c=\psi_c$. The result for $\psi''_{c+1}$ can be shown by simply following identical logic.
 \item[(iv)] From Equation \ref{eqn:stick}, we have
\begin{eqnarray}
\nonumber
 \mathbb{E}[\psi_c|\bZ,\alpha] & = & \mathbb{E}[V_c\prod_{l<c}(1-V_l)|\bZ,\alpha]\\
\nonumber
& = & \mathbb{E}[V_c|\bZ,\alpha]\prod_{l<c}\mathbb{E}[(1-V_l)|\bZ,\alpha]\\
\label{eqn:epsicz}
& = & \left(\frac{1+n_c}{1+\alpha+n_c+\sum_{l>c}n_l}\right)\\
&&\times \prod_{l<c}\left(\frac{\alpha+\sum_{l'>l}n_{l'}}{1+\alpha+n_l+\sum_{l'>l}n_{l'}}\right).
\end{eqnarray}
Similarly,
\begin{eqnarray}
\nonumber
  \mathbb{E}[\psi_{c+1}|\bZ,\alpha]  & = & \left(\frac{1+n_{c+1}}{1+\alpha+n_{c+1}+\sum_{l>c+1}n_l}\right)\\
\label{eqn:epsicplus1z}
& &\times \left(\frac{\alpha+\sum_{l>c}n_{l}}{1+\alpha+n_c+\sum_{l>c}n_{l}}\right)\\
\nonumber
& & \times \prod_{l<c}\left(\frac{\alpha+\sum_{l'>l}n_{l'}}{1+\alpha+n_l+\sum_{l'>l}n_{l'}}\right).
\end{eqnarray}
By definition of $\bZ'$ in Equation \ref{eqn:labelswitchZ}, we have
\begin{equation}
\label{eqn:ndash}
 n'_l=\begin{cases}
	n_{c+1} & l=c\\
	n_c & l=c+1\\
	n_l & \textrm{otherwise.}
     \end{cases}
\end{equation}

This means from Equations \ref{eqn:epsicz} and \ref{eqn:epsicplus1z} we have
\begin{eqnarray}
\nonumber
 \frac{\mathbb{E}[\psi_c|\bZ',\alpha]}{\mathbb{E}[\psi_{c+1}|\bZ,\alpha]} & = & \left(\frac{1+n'_c}{1+\alpha+n'_c+n'_{c+1}+\sum_{l>c+1}n_l}\right)\\
\label{eqn:epsiratio}
   & & \times \left(\frac{1+\alpha+n_{c+1}+\sum_{l>c+1}n_l}{1+n_{c+1}}\right)\\
\nonumber
   & & \times \left(\frac{1+\alpha+n_c+n_{c+1}+\sum_{l>c+1}n_l}{\alpha+n_{c+1}+\sum_{l>c+1}n_l}\right)
\end{eqnarray}
Substituting Equation \ref{eqn:ndash} into \ref{eqn:epsiratio} and simplifying gives the desired results. The result for
$\frac{\mathbb{E}[\psi_{c+1}|\bZ',\alpha]}{\mathbb{E}[\psi_c|\bZ,\alpha]}$ follows in the same fashion.

\item[(v)] By (iii) and the deterministic nature of the proposal mechanism, the only random feature of the proposal is the choice of component $c$. The probability of this choice is the same for the move and its reverse and so cancels. Therefore the only contribution to the acceptance ratio is the ratio of posteriors. By design, the likelihood is unchanged, and by (ii) the only change in posterior is down to the change in weights of components $c$ and $c+1$. Therefore we have,
\begin{eqnarray}
 R &=& \frac{(\psi'_{c})^{n'_c}(\psi'_{c+1})^{n'_{c+1}}}{\psi_c^{n_c}\psi_{c+1}^{n_{c+1}}}\\ &=& \left(\frac{\psi'_{c+1}}{\psi_c}\right)^{n_c}\left(\frac{\psi'_c}{\psi_{c+1}}\right)^{n_{c+1}} \;\;\; \textrm{by Equation \ref{eqn:ndash}}.
\end{eqnarray}
Substituting in Equation \ref{eqn:labelswitchPsi} and the results in (iv), we obtain the desired acceptance ratio.
\end{enumerate}
\end{proof}
\end{proposition}
\end{appendix}

\bibliographystyle{spbasic}
\bibliography{dirichletSamplerPaper_2013}

\end{document}


\maketitle

\section{Simulated data}
Here we provide additional details describing how we simulated the datasets that were used to illustrate our work.
\subsection{ Dataset 1 - used for paper Figure 1}
This dataset is a simple simulation using aspects of a profile regression model, conditioning on 1000 observations being split into 5 clusters, each containing 200 observations. We assume a Bernoulli response model and 10 discrete covariates, each with 2 categories. No fixed effects or missing data points were included. 

For each cluster $c$, the response and covariate data for the 200 observations were generated by sampling from the model using the the response parameter $\theta_c$ and the covariate model parameters $\Phi_c$ in Table \ref{tab:tab1} below. The values are fixed to ensure that the clusters were well separated within the parameter space.

\begin{table}[h]
\begin{center}
\begin{tabular}{cccccc}
&Cluster 1&Cluster 2&Cluster 3&Cluster 4&Cluster 5\\
\hline
\hline
$\theta_c$&-2.19&-0.84&0&0.84&2.19\\
$\mathbb{P}(Y=1) $&0.1&0.3&0.5&0.7&0.9\\
\hline
$\mathbb{P}(X_1=0)$&0.9&0.9&0.1&0.1&0.1\\
$\mathbb{P}(X_2=0)$&0.9&0.9&0.9&0.1&0.1\\
$\mathbb{P}(X_3=0)$&0.9&0.9&0.1&0.1&0.1\\
$\mathbb{P}(X_4=0)$&0.9&0.9&0.9&0.1&0.1\\
$\mathbb{P}(X_5=0)$&0.1&0.9&0.1&0.1&0.9\\
$\mathbb{P}(X_6=0)$&0.1&0.9&0.9&0.1&0.9\\
$\mathbb{P}(X_7=0)$&0.1&0.9&0.1&0.1&0.9\\
$\mathbb{P}(X_8=0)$&0.1&0.9&0.9&0.1&0.9\\
$\mathbb{P}(X_9=0)$&0.9&0.9&0.1&0.9&0.9\\
$\mathbb{P}(X_{10}=0)$&0.1&0.1&0.9&0.1&0.1\\
\end{tabular}
\caption{Response and covariate parameters used to generate simulated dataset 1.\label{tab:tab1}}
\end{center}
\end{table}

\clearpage
\section{Dataset 2 - used for paper Figure 9}
The second dataset is generated more faithfully from a profile regression model, allowing the cluster allocations and cluster parameters to be generated according to their prior distributions. We simulate $n=2,000$ observations. Each individual $i$ has a discrete covariate vector $X_i$ of $J=10$ covariates (with each covariate sampled from one of $K_j=5$ categories), a vector $W_i$ of $L=10$ fixed effects, and a Bernoulli outcome $Y_i$.

We use the following algorithm to generate the data.

\begin{itemize}
\item[A]
\begin{itemize}
\item[A.1] Set $C^{\star}=1$, $C=1$, $i=0$, and $\psi_1 = 0$
\item[A.2] If $i=n$ go to B
\item[A.3] Set $i=i+1$
\item[A.4] Simulate $u\sim\mathrm{Unif}(0,1)$
\item[A.5] If $u < \sum_{c<=C}\psi_c$ set $z_i = C$, $i=i+1$, $C=1$ and go to A.2
\item[A.6] If $C<C^{\star}$ set $C=C+1$ and go to A.5
\item[A.7] Simulate $\alpha_{C^{\star}} \sim \mbox{Gamma}(\mbox{shape}=9,\mbox{scale}=0.5)$
\item[A.8] Simulate $v_{C^{\star}} \sim \mathrm{Beta}(1, \alpha_{C^{\star}})$
\item[A.9] Compute $\psi_{C^{\star}} = v_{C^{\star}} \prod_{l<C^{\star}} (1-v_l)$
\item[A.10] Set $C=C+1$, $C^{\star}=C^{\star}+1$ and go to A.5
\end{itemize}
\item[B] For $c=1,\ldots,C^{\star}$, we generate $\theta_c \sim t_7(0, 1)$ for the outcome and $\phi_{c,j} \sim \mbox{Dirichlet}(1,1,1,1,1)$ for $j=1,\ldots,J$ for the covariates 
\item[C] For $l=1,\ldots,L$ generate $\beta_l \sim t_7(0,1)$ for the fixed effects coefficients. 
\item[D] For $i=1,\ldots,n$ generate the fixed effect data $W_i$ from a $\mbox{Normal}_L(0,I_L)$
\item[E] For $i=1,\ldots,n$ compute $\lambda_i = \theta_{z_i} + \beta W_i$.
\item[F] For $i=1,\ldots,n$ generate $Y_i$ where $\mathbb{P}(Y_i=1)=\mbox{logit}^{-1}(\lambda_i)$.
\item[G] For $i=1,\ldots,n$ generate $X_{i,j}$ according to the probabilities $\phi_{z_i,j}$, for each $j=1,\ldots,J$.
\end{itemize}

\section{Computation of marginal partition posterior}
Here we provide further details regarding the computation of the marginal partition posterior, defined in Section 4.1 of the paper. To illustrate the computation we use the profile regression model used in the paper with a Bernoulli response and discrete categorical covariates. 

Most generally, the marginal partition posterior is defined as $p(\bZ|\bD)$.

For the case of profile regression, our data $\bD=(\bY,\bX)$ is split into the response data $\bY$ and the covariate data $\bX$. We must also condition on the fixed effect data $\bW$. We can write 
\[
p(\bZ|\bD, \bW) = p(\bZ|\bY, \bX, \bW) \propto p(\bX|\bZ)p(\bY|\bZ,\bW)p(\bZ). 
\]
The three factors on the right hand side are (from left to right) the likelihood of the covariate data, the likelihood of the response data and the prior for the partition. To proceed we look at each of these in turn.

\subsection{Likelikood of covariate data}
First we consider $p(\bX|\bZ)$. Clearly computation of this factor depends entirely upon the covariate model. In the case considered in this paper the covariate model is a discrete categorical model, meaning we can write 
\[
p(\bX|\bZ) = \int p(\bX| \Phi, \bZ) p(\Phi) \mbox{d}\Phi.
\]

This gives
\[p(\bX|\bZ) = \int \left[ \prod_{i=1}^n \prod_{j=1}^J \phi_{z_i,j,x_{ij}} \right] \prod_{c=1}^\infty \prod_{j=1}^J \left[ \frac{\Gamma(K_j a)}{\prod_{k=1}^{K_j} \Gamma(a)} \prod_{k=1}^{K_j} \phi_{c,j,k}^{a-1} \right] \mbox{d}\Phi
\]

where for $c=1,2,\ldots$, 
\[
p(\phi_{c,j}) \sim \mbox{Dirichlet} (\underbrace{a,\ldots,a}_{K_j \mbox{ times}})
\]
and $K_j$ is the number of categories for covariate $j$.

For empty clusters the integral is simply the integral of $\Phi_c$ over its prior. Since there is can only be a finite number of non-empty clusters, the infinite product can be written as a finite product meaning
\[p(\bX|\bZ) = \int \left[ \prod_{i=1}^n \prod_{j=1}^J \phi_{z_i,j,x_{ij}} \right] \prod_{c:n_c>0} \prod_{j=1}^J \left[ \frac{\Gamma(K_j a)}{\prod_{k=1}^{K_j} \Gamma(a)} \prod_{k=1}^{K_j} \phi_{c,j,k}^{a-1} \right] \mbox{d}\Phi,
\]
 where $n_c$ is the number of observations in cluster $c$.

This is the integral over a finite product of unnormalised Dirichlet distributions, giving 
\[
p(\bX|\bZ) =  \prod_{c:n_c > 0} \prod_{j=1}^J \left[ \frac{\Gamma(K_j a)}{\prod_{k=1}^{K_j} \Gamma(a)} \frac{\prod_{k=1}^{K_j} \Gamma(a+n_{c,j,k})}{\Gamma(K_j a + n_c)} \right] 
\]
where $n_{c,j,k}= \sum_i 1_{\{z_i = c\} \cap \{x_{ij}=k\}}$. 

\subsection{Likelikood of response data}
Next we consider the $p(\bY|\bZ,\bW)$. For the model considered in the paper, we assume a Bernoulli response, so that
\begin{eqnarray}
\nonumber  
p(\bY|\bZ,\bW) &=& \int p(\bY|\theta,\beta,\bZ,\bW)p(\theta)p(\beta)\mbox{d}\theta \mbox{d}\beta \\
\nonumber
&=&\int \prod_{i=1}^n \left(  \frac{e^{\theta_{z_i}+\beta^T\bW}}{1+e^{\theta_{z_i}+\beta^T\bW}} \right)^{Y_i}\left(  \frac{1}{1+e^{\theta_{z_i}+\beta^T\bW}} \right)^{1-Y_i}p(\theta)p(\beta)\mbox{d}\theta \mbox{d}\beta
\end{eqnarray}

This is not a standard distribution so we use a multivariate Laplace approximation to approximate the integral. For a $d$-vector $\eta$, a function $h:\mathbb{R}^d\rightarrow\mathbb{R}$, and a large number $M$ the multivariate Laplace approximation can be written as

\[
\int e^{-Mh(\eta)}\mbox{d} \eta \approx  e^{-Mh(\hat\eta)} |\Sigma|^{\frac{1}{2}} M^{\frac{d}{2}}(2\pi)^{\frac{1}{2}}
\]
where $\hat \eta$ is the global minimum of $h$ and $\Sigma$ is the inverse Hessian of $h$ evaluated at $\hat \eta$. Noting that we only need $p(\bY|\bZ,\bW)$ up to a constant, we can discard the final factor of the right hand side. Then, defining $M=n$ and $\eta = (\theta^\star, \beta)$, where $\theta^\star$ contains the elements of $\theta$ corresponding to non-empty clusters (the other $\theta$'s are just integrated over the prior) we obtain

\begin{equation}
p(\bY|\bZ,\bW) \propto  e^{-nh(\hat\eta)} |\Sigma|^{1/2} n^{-\frac{1}{2}(C^\star+L)}
\label{py}
\end{equation}
where $C^\star$ is the number of non-empty clusters and $L$ is the number of fixed effects. The function $h$ is given by
\[
h(\eta) = -\frac{1}{n}\left( \sum_{i=1}^n \left[ Y_i(\theta_{z_i}+\beta^T W_i)-\log(1+e^{\theta_{z_i}+\beta^T W_i})\right]+\log p(\theta^\star)+\log p(\beta)\right) 
\]
with 
\[
\log p(\theta^\star) = - \sum_{c: n_c > 0}^C \frac{\nu+1}{2} \log \frac{\nu+\frac{\theta_c^2}{\sigma_\theta^2}}{\nu} + \mbox{constants}
\]
and
\[
\log p(\beta) = - \sum_{l=1}^L \frac{\nu+1}{2} \log \frac{\nu+\frac{\beta_l^2}{\sigma_\beta^2}}{\nu}+ \mbox{constants},
\]
where $\nu$ is the number of degrees of freedom of the t-distributions for $\theta_c$ and $\beta_l$. 

To compute the exact value we thus proceed by 
\begin{enumerate}
\item Finding $\hat\eta$, the value of $\eta$ that minimises $h$
\item Evaluating $h$ and the Hessian of $h$ at $\hat\eta$
\item Plugging these into Equation (\ref{py})
\end{enumerate}

\subsection{Prior of partition}
The final factor is the prior distribution of the partition $p(\bZ)$. This can be written as:
\[
p(\bZ) = \int p(\bZ | \bV) p(\bV| \alpha) \mbox{d} \bV \mbox{d}\alpha. 
\]

In order to simplify this calculation, we assume a fixed value of $\alpha = \alpha^\star$, which in our case was chosen as the posterior mean of our MCMC sample. Thus we can write
\begin{eqnarray}
\nonumber  
p(\bZ) &=& \int p(\bZ | \bV) p(\bV| \alpha^\star) \mbox{d} \bV\\
\nonumber
&=& \int \prod_{i=1}^n \left[V_{z_i} \prod_{c < z_i} (1-V_c) \right] \prod_{c=1}^\infty \frac{1}{B(1,\alpha^\star)} (1-V_c)^{\alpha^\star-1} \mbox{d}\bV,
\end{eqnarray}
where $B(a,b)$ denotes the Beta function.

Defining $C =\max_i z_i$, then
\[
p(\bZ) = \int \prod_{i=1}^n \left[V_{z_i} \prod_{c < z_i} (1-V_c) \right] \prod_{c=1}^C \frac{1}{B(1,\alpha^\star)} (1-V_c)^{\alpha^\star-1} \mbox{d}V_1 \mbox{d}V_2 \ldots \mbox{d}V_C
\]
since for the other empty clusters we are integrating $V$ over the prior. 

Following Lijoi et al. (2008), it follows that 
\[
p(\bZ) = \frac{n!\Gamma(\alpha^\star)}{\Gamma(\alpha^\star+n)}\prod_{j=1}^n \frac{(\alpha^\star)^{a_j}}{j^{a_j} a_j!}
\]
where $a_j = \#\{c: n_c = j\}$. 

\section*{Additional References}
Lijoi A, Pruenster I, Walker SG. (2008). Bayesian nonparametric estimators derived from conditional Gibbs structures. The Annals of Applied Probability, 18, 1519-1547